\documentclass[twocolumn,showpacs,showkeys,preprintnumbers,amsmath,amssymb]{revtex4}

\usepackage{graphicx}
\usepackage{dcolumn}
\usepackage{bm}

\begin{document}

\title{A method for the quantitative study of atomic transitions in a magnetic field
based on an atomic vapor cell with L=$\lambda $}

\author{Armen Sargsyan, Grant Hakhumyan, Aram Papoyan, David Sarkisyan$^*$}
\affiliation{Institute for Physical  Research, Armenian Academy of Sciences, Ashtarak-0203, Armenia}

\author{Aigars Atvars, and Marcis Auzinsh}                      
\affiliation{Department of Physics, University of Latvia, 19 Rainis blvd., Riga, LV-1586 Latvia}

\date{\today}

\begin{abstract}

We describe the so-called "$\lambda$-Zeeman method" to investigate individual hyperfine transitions
between Zeeman sublevels of atoms in an external magnetic field of 0.1 mT $\div$ 0.25 T. Atoms are confined
in a nanocell with thickness \textit{L} = $\lambda$, where $\lambda$ is the resonant wavelength (794 nm or 780
nm for D$_{1}$ or D$_{2}$ line of Rb). Narrow resonances in the transmission spectrum of the nanocell are split into
several components in a magnetic field; their frequency positions and probabilities depend on the \textit{B}-field.
Possible applications are described, such as magnetometers with nanometric spatial resolution and tunable atomic
frequency references.
\end{abstract}


\maketitle

\indent It is well known that energy levels of atoms placed in an external magnetic field undergo frequency shifts
and changes in their transition probabilities. These effects were studied for hyperfine (hf) atomic transitions
in the transmission spectra obtained with an ordinary cm-size cell containing Rb and Cs vapor was used in \cite{Ref1}.
However, because of Doppler-broadening (hundreds of MHz), it was possible to partially separate different hf transitions
only for \textit{B} $>$ 0.15 T. Note that even for these large \textit{B} values the lines of $^{87}$Rb and $^{85}$Rb are strongly
overlapped, and pure isotopes have to be used to avoid complicated spectra. In order to eliminate the Doppler broadening,
the well-known saturation absorption (SA) technique was implemented to study the Rb hf transitions \cite{Ref2}. However, in this
case the complexity of the Zeeman spectra in a magnetic field arises primarily from the presence of strong crossover (CO)
resonances, which are also split into many components. That is why, as is mentioned in \cite{Ref2}, the SA technique is
applicable
only for \textit{B} $<$ 5 mT. The CO resonances can be eliminated with selective reflection (SR) spectroscopy \cite{Ref3}, but to
correctly determine of the hf transition position, the spectra must undergo further non-trivial processing. Another method
based on the fluorescence spectrum emitted from a nanocell at thickness \textit{L} = $ \lambda $/2 was presented in \cite{Ref4}.
However, in this case the sub-Doppler spectral line-width is relatively large ($\sim $ 100 MHz); also the laser power has to
be relatively large to detect a weak fluorescence signal. Coherent population trapping (CPT) allows one to study the
behavior of hf transitions in a magnetic field with very high accuracy (several kHz) \cite{Ref5}, however the experimental
realization is complicated; moreover, measuring hf level shifts of several GHz for \textit{B} $\sim ${} 0.1 T using CPT
is not realistic.
\begin{figure}[hbtp]
 \centering
 \includegraphics [scale=0.98]{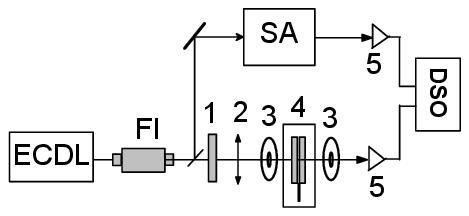}
 \caption{Experimental setup. FI - Faraday isolator,
 1 - $\lambda $/4 plate, 2 - lens (\textit{F} = 35 cm),
 3 - ring magnets,
4 - nanocell and the oven,
5 - photodetectors, DSO-digital storage oscilloscope.}
 \end{figure} \\
 \indent We present a method based on narrow (close to natural linewidth)
velocity selective optical pumping/saturation (VSOP) resonance peaks of reduced absorption located at the atomic transitions \cite{Ref6}.
The VSOP peaks appear at laser intensity
$\sim$ 1 mW/cm$^{2}$ in the transmission spectrum of the nanocell with atomic vapor column of thickness\textit{ L} = $\lambda$,
where $\lambda$ is the resonant wavelength of the laser radiation (794 nm or 780 nm for Rb D$_{1}$ or D$_{2}$ line). At \textit{B} $>$ 0,
the VSOP resonance is split into several Zeeman components, the number of which depends on the quantum numbers \textit{F} of
the lower and upper levels. The amplitudes of these peaks and their frequency positions depend unambiguously on the \textit{B}
value. This so called "$\lambda$-Zeeman method" (LZM) allows one to study not only the frequency shift of any individual hf
transition, but also the modification in transition probability in the region of 0.1 mT - 0.25 T (LZM is expected to be valid
up to several T).
\begin{figure}[hbtp]
 \centering
 \includegraphics [scale=0.98]{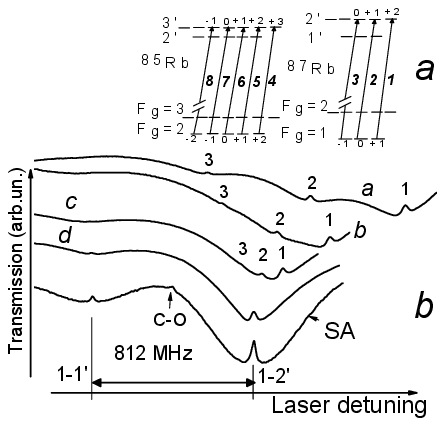}
 \caption{ a) $^{87}$Rb, $^{85}$Rb D$_{1}$ line atomic transitions,
 $\sigma^{+}$ excitation; b)\textit{ F$_{g}$ }= 1 $\rightarrow$ \textit{F$_{e}$} = 1,2
transmission spectra for B = 59 mT (a), 31 mT (b),
11.5 mT (c), and 0 (d); lower curve is SA spectrum.}
 \end{figure} \\
\indent Experimental realization of LZM is simple enough (see Fig.1). The circularly polarized beam of an extended cavity diode
laser (ECDL, $\lambda$ = 794 nm, \textit{P$_{L}$} $\sim $ 5 mW, $\gamma_{L} $ $<$ 1 MHz) resonant with the $^{87}$Rb D$_{1}$ transition
frequency, after passing through Faraday isolator, was focused ($\oslash$ 0.5 mm) onto Rb nanocell with a vapor column of
 thickness \textit{L} = $\lambda$ at an angle close to normal. The design of a nanocell is presented in \cite{Ref6}. The
source temperature of the atoms of the nanocell was 110 $^{\circ}$C, corresponding to a vapor density \textit{N} $\sim$
$10^{13}$ cm$^{3}$, but the windows were maintained at a temperature that was 20 $^{\circ}$C higher. Part of the laser radiation
was diverted to a cm-size Rb cell to obtain a \textit{B} = 0 SA spectrum, which served as frequency reference.
The nanocell transmission and SA spectra were detected by photodetectors and recorded by a digital storage oscilloscope.
\begin{figure}[hbtp]
 \centering
 \includegraphics [scale=0.98]{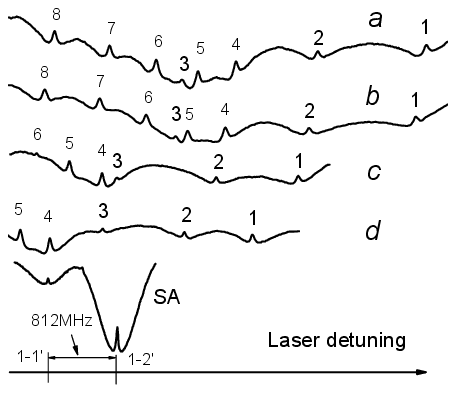}
 \caption{ Transmission spectra for B = 0.24 T (a), 0.23 T (b), 0.154 T (c), 0.117 T (d) ;
 lower curve is SA spectrum.}
 \end{figure}
 Small longitudinal magnetic fields (\textit{B} $<$ 25 mT) were applied to the nanocell by a system of Helmholtz coils
(not shown in Fig.1). The \textit{B}-field strength was measured by a calibrated Hall gauge. Among the advantages of LZM
is the possibility to apply much stronger magnetic fields using widely available strong permanent ring magnets (PRM).
In spite of the strong inhomogeneity of the \textit{B}-field (in our case it can reach 15 mT/mm), the variation of \textit{B}
inside the atomic vapor column is a few $\muµ$T, i.e., by several orders less than the applied \textit{B} value because of the
small thickness of the nanocell (794 nm). The allowed transitions between magnetic sublevels of hf states for
the $^{87}$Rb D$_{1}$
line in the case of $\sigma^{+} $ (left circular) polarized excitation are depicted in Fig. 2a (LZM also works well
for $\sigma^{-} $ excitation). Fig. 2b shows the nanocell transmission spectra for the \textit{F$_{g}$}=1 $\rightarrow$ \textit{F$_{e}$}=1,2
transitions at different values of \textit{B} (the labels denote corresponding transitions shown in Fig.2a). As it is seen,
all the individual
Zeeman transitions are clearly detected. The two transitions\textit{ F$_{g}$}=1 $\rightarrow$ \textit{F$_{e}$}=1 (not shown in Fig. 1a)
are detectable for \textit{B} $<$ 12 mT, while at higher \textit{B} their probabilities are strongly reduced (this is also
confirmed theoretically \cite{Ref4}).  Note that the absence of CO resonances in transmission spectra is an important advantage
 of the nanocell \cite{Ref6}.
\begin{figure}[hbtp]
 \centering
 \includegraphics [scale=0.8]{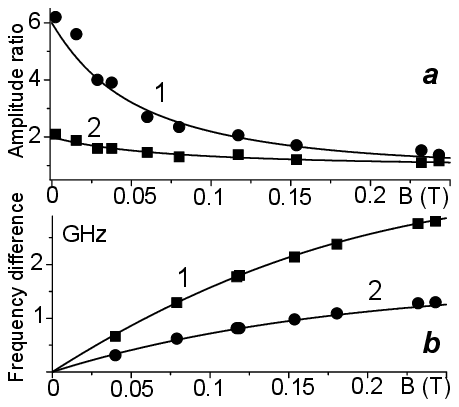}
 \caption{ a) (1) - ratio A(1)/A(3), (2) - ratio A(1)/A(2) versus B; b) (1) - $\Delta$(1,3), (2)- $\Delta$(1,2) versus B.}
 \end{figure} \\
\indent Transmission spectra for larger \textit{B} values are presented in Fig. 3 (also for Zeeman transitions of the $^{85}$Rb D$_{1}$ line).
The strong magnetic field was produced by two $\oslash$ 30 mm PRMs with $\oslash$ 3 mm holes to allow radiation to pass
placed on opposite sides of the nanocell oven and separated by a distance that was varied between 35 and 50 mm (see Fig. 1).
To control the magnetic field value, one of the magnets was mounted on a micrometric translation stage for longitudinal
displacement. In particular, the \textit{B}-field difference of curves (\textit{a}) and (\textit{b}) is obtained by a PRM
displacement of 0.67 mm, corresponding to a rate of 15 mT/mm. The frequency difference between the VSOP peaks
numbered\textit{ 4 }(curves (\textit{a})\textit{,} (\textit{b})) for this case is 100 MHz. By a 20 $\mu$m displacement of
the PRM it is easy to detect $\sim $ 3 MHz frequency shift of peak \textit{4}. The advantage of submicron-size magnetic
field probe can be fully exploited for the case of larger \textit{B}-field gradient, as well as after further optimization of
the method (reduction of laser intensity, implementation of frequency modulation and lock-in detection, etc.).
An important advantage of LZM is that the amplitude of VSOP peaks is linearly proportional to the corresponding Zeeman
transition probability, which offers the possibility to quantitatively study the modification of individual Zeeman
transition probabilities in a magnetic field. Thus, in weak magnetic fields (\textit{B} $\sim$  0), the probabilities of
 transitions labeled \textit{1}, \textit{2}, and \textit{3} compose the ratio 6:3:1, which varies rapidly as \textit{B}
increases. Fig. 4a presents the amplitude ratio A(\textit{1})/A(\textit{3}) (curve 1) and A(\textit{1})/A(\textit{2})
(curve 2) as a function of \textit{B} (hereafter the dots and solid lines denote experiment and theory, respectively).
Fig. 4b shows the frequency difference $\Delta$(\textit{1,3}) and $\Delta$(\textit{1,2}) between transitions
labeled \textit{1} and \textit{3} (curve 1) and \textit{1} and 2 (curve 2) as a function of \textit{B}. Obviously,
by measuring $\Delta$(\textit{1,3}) and $\Delta$(\textit{1,2}) it is possible to determine the strength of magnetic
field, even in the absence of reference spectra.
\begin{figure}[hbtp]
 \centering
 \includegraphics [scale=0.8]{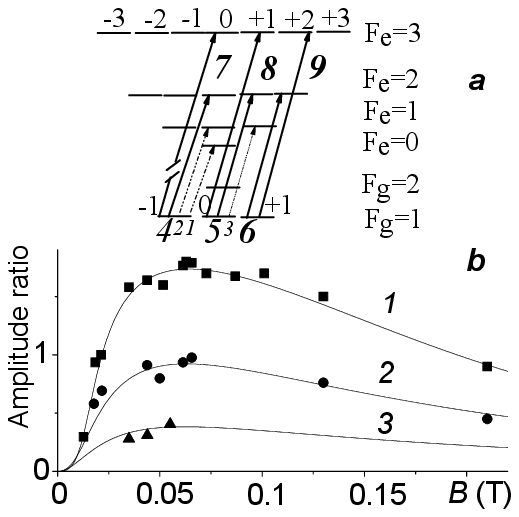}
 \caption{ a) $^{87}$Rb D$_{2}$ line atomic transitions, $\sigma^{+}$ excitation;
 b) (1) - ratio A(7)/A(6), (2) - ratio A(8)/A(6), (3) - ratio A(9)/A(6) versus B.}
 \end{figure}
 \\
\indent We also implemented LZM to study transitions\textbf{ }\textit{Fg}=1 $\rightarrow$ \textit{Fe}=0,1,2,3 of the $^{87}$Rb D$_{2}$
line ($\lambda$ = 780 nm; all the other experimental parameters and conditions are the same). The possible Zeeman
transitions for $\sigma^{+}$ polarized excitation are depicted in Fig. 5a. Particularly, it was revealed that for
 \textit{B }$>$ 10 mT, also the $^{87}$Rb D$_{2}$, \textit{F$_{g}$}=1 $\rightarrow$ \textit{F$_{e}$}=3 "forbidden" transitions
 (labeled \textit{7,8,9}) appear in spectrum, for which new selection rules with respect to the quantum number
\textit{F} apply. Moreover, for 20 mT $<$ \textit{B} $<$ 0.2 T the probability of transition \textit{7} exceeds that
of \textit{6}, the strongest
transition at \textit{B} = 0. Fig. 5b gives the \textit{B}-field dependence of amplitude (transition probability)
ratio A(\textit{7})/A(\textit{6}), A(\textit{8})/A(\textit{6}), and A(\textit{9})/A(\textit{6}) (curves 1,2,3).
Both in Fig. 4a and Fig. 5b, there is good agreement between experiment and theory. The nanocell transmission spectrum
for these transitions at \textit{B} = 0.21 T is presented in Fig. 6. The two arrows show the positions of the
$^{85}$Rb \textit{F$_{g}$}=2 $\rightarrow$ \textit{F$_{e}$}=4 Zeeman transitions (theory \cite{Ref4} well predicts that their probabilities
have to be small).
\begin{figure}[hbtp]
 \centering
 \includegraphics [scale=0.8]{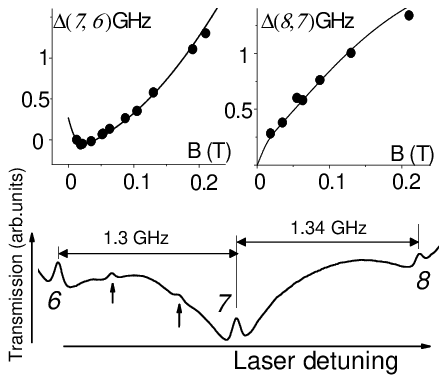}
 \caption{ Transmission spectrum, B = 0.21 T; upper insets: $\Delta$(7,6) (left), and  $\Delta$(8,7) (right) versus B.}
 \end{figure}
The upper insets show the \textit{B}-field dependence of $\Delta$(\textit{7,6}) and $\Delta$(\textit{8,7}).
We should note that transition \textit{7} is strongly shifted (by 5.6 GHz) from the \textit{B} = 0 position
of the \textit{F$_{g}$}=1 $\rightarrow$ \textit{F$_{e}$}=2 transition.
The latter allows development of a frequency reference
 based on a nanocell and PRMs, widely tunable over a range of several GHz by simple displacement of the magnet. LZT can
be successfully implemented also for studies of the D$_{1}$ and D$_{2}$ lines of Na, K, Cs, and other atoms.

\section*{Acknowledgements}
This work is partially supported by INTAS South-Caucasus Grant 06-1000017-9001 and by SCOPES Grant IB7320-110684/1.
We acknowledge support from the ERAF grant VPD1/ERAF/CFLA/05/APK/2.5.1./000035/018, and A.A acknowledges support from
the ESF project.

E-mail of David Sarkisyan:  david@ipr.sci.am.


\end{document}